# Strong Enhancement of Spin-Orbit Torques and Perpendicular Magnetic Anisotropy in [Pt$_{0.75}$Ti$_{0.25}$/Co-Ni multilayer/Ta]$_n$ Superlattices

Xiaomiao Yin[1,2,3], Zhengxiao Li[2,4], Jun Kang[1,3*], Changmin Xiong[3,5], Lijun Zhu[2,4**]

1. Beijing Computational Science Research Center, Beijing, 100193, China
2. State Key Laboratory of Semiconductor Physics and Chip Technologies, Institute of Semiconductors, Chinese Academy of Sciences, Beijing 100083, China
3. Department of Physics, Beijing Normal University, Beijing 100875, China
4. Center of Materials Science and Optoelectronics Engineering, University of Chinese Academy of Sciences, Beijing 100049, China
5. Key Laboratory of Multiscale Spin Physics, Ministry of Education, Beijing Normal University, Beijing 100875, China

*jkang@csrc.ac.cn, **ljzhu@semi.ac.cn

**Abstract:** We report the development of the [Pt$_{0.75}$Ti$_{0.25}$/Co-Ni multilayer/Ta]$_n$ superlattice with strong spin-orbit torque, large perpendicular magnetic anisotropy, and low switching current density. We demonstrate that the efficiency of the spin-orbit torque increases linearly with the repetition number $n$, which is in good agreement with the spin Hall effect of the Pt$_{0.75}$Ti$_{0.25}$ being the only source of the spin-orbit torque. Meanwhile, the perpendicular magnetic anisotropy field is also enhanced by more than a factor of 2 as $n$ increases from 1 to 6. The [Pt$_{0.75}$Ti$_{0.25}$/(Co/Ni)$_3$/Ta]$_n$ superlattice also exhibits deterministic, low-current-density magnetization switching despite the very large layer thicknesses. The combination of the strong spin-orbit torque, perpendicular magnetic anisotropy, and low-current-density switching makes the [Pt$_{0.75}$Ti$_{0.25}$/Co-Ni multilayer/Ta]$_n$ superlattice a compelling material candidate for ultrafast, energy-efficient, long-data-retention spintronic technologies.

It has been a central topic to develop magnetic heterostructures that simultaneously have strong perpendicular magnetic anisotropy, spin-orbit torques (SOTs), and efficient current switching for spin-based memory, computing, and sensor technologies [1–10]. The efficiencies of the SOTs are directly relevant to the switching current and thus energy efficiency of the devices in practical device applications [1,11,12]. Meanwhile, the perpendicular magnetic anisotropy (PMA) determines the thermal stability and data retention of the nanodevices fabricated from the magnetic heterostructures. Lowering the device dimension for high-density integration typically brings a beneficial reduction of the SOT switching current and detrimental loss of thermal stability and data retention at the same time [2]. Thus, the development of strategies that can enhance the PMA and the SOT of spin-orbit heterostructures is of great importance. Efforts in the past decade have established that SOTs can be considerably increased by a moderate increase of resistivity of a spin Hall metal by alloying [13-19] or atomic insertions [20,21] or by stacking symmetry-broken spin-orbit superlattices [22,23]. However, there has been no report on spin-orbit superlattices that combine a strong spin Hall alloy and strong PMA.

In this work, we, for the first time, report the development of the [Pt$_{0.75}$Ti$_{0.25}$/Co-Ni multilayers/Ta]$_n$ superlattices with enhanced SOTs, PMA, and low-current-density switching by combining the strong spin Hall alloy of Pt$_{0.75}$Ti$_{0.25}$ [24] and the PMA Co-Ni multilayer. The Co-Ni multilayer is chosen as the ferromagnet since it has been suggested by previous experiments to have tunable PMA [25-29], low magnetization, small magnetic damping [30], ultrafast domain wall motion and switching dynamics [31-35], and large tunnel magnetoresistance [36].

As shown in Fig. 1(a), we sputter-deposited five spin-orbit superlattices of [Pt$_{0.75}$Ti$_{0.25}$ (2 nm)/Co-Ni multilayer/Ta (1.5 nm)]$_n$ with different repeat numbers $n$ of 1, 3, 4, 5, and 6 on the oxidized Si substrates at room temperature. For each repeat of the superlattice, the Co-Ni multilayer consists of [Co (0.29 nm)/Ni (0.36 nm)]$_3$ and functions as the spin current detector, the Pt$_{0.75}$Ti$_{0.25}$ is the spin current generator, while the 1.5 nm disordered Ta grown with a low rate of 0.0092 nm/s is chosen following the previous in-plane magnetized superlattice [23] to break the symmetry by blocking the spin current of the neighboring Pt$_{0.75}$Ti$_{0.25}$ from flowing into the Co-Ni multilayer underneath the Ta (Fig. 1(a)) but generates negligible spin current (Fig. 1(b)) likely due to the small thickness, very high resistivity [13-23], and disordered structure due to the low growth rate [37]. Note that, without the Ta symmetry-breaking layer, no SOTs would be expected on the Co-Ni multilayers because the spin currents from the two Pt$_{0.75}$Ti$_{0.25}$ layers below and above the very Co-Ni multilayers would fully cancel out (Fig. 1(b)). The subscripts of the Pt$_{0.75}$Ti$_{0.25}$ represent the volume percentages as calibrated using the deposition rates of the Pt and the Ti. Each superlattice is seeded by 1 nm Ta for improved adhesion and protected from oxidation by a top MgO (1.6 nm)/Ta (1.5 nm) bilayer, forming a total stack of Si/SiO$_2$/Ta (1 nm)/superlattice/MgO (1.6 nm)/Ta (1.5 nm). As shown in Fig. 1(c), the spin-orbit superlattices were patterned into 5×60 $\mu$m$^2$ Hall bars by photolithography and argon ion etching, followed by the deposition of Ti (5 nm)/Pt (150 nm). All the Hall bar devices are annealed at 250°C and zero magnetic field in a vacuum of 1×10$^{-9}$ Torr for enhanced PMA.

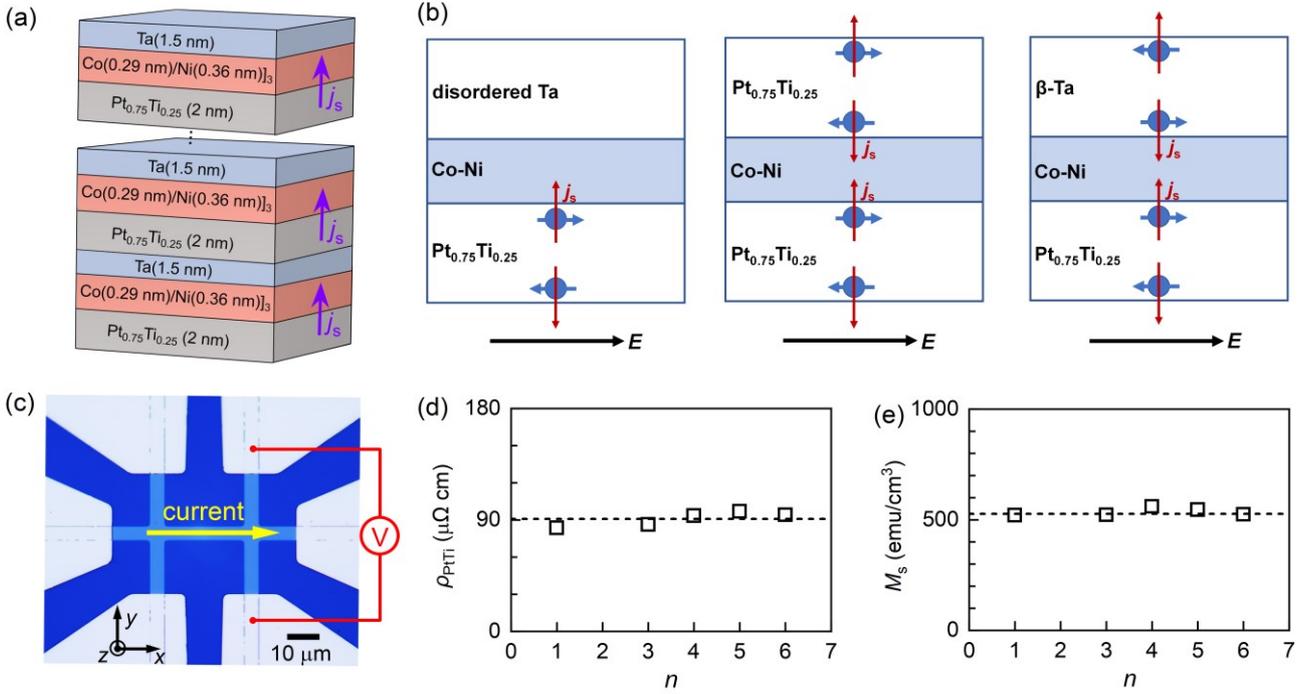

Fig. 1. Sample characteristics. (a) Schematic of the superlattice. (b) Spin current configurations in disordered Ta/Co-Ni multilayer/$Pt_{0.75}Ti_{0.25}$, $Pt_{0.75}Ti_{0.25}$/Co-Ni multilayer/$Pt_{0.75}Ti_{0.25}$, and β-Ta/Co-Ni multilayer/$Pt_{0.75}Ti_{0.25}$. Spin currents from the under- and over-layers of the Co-Ni multilayer cancel out in $Pt_{0.75}Ti_{0.25}$/Co-Ni multilayer/$Pt_{0.75}Ti_{0.25}$ and add in β-Ta/Co-Ni multilayer/$Pt_{0.75}Ti_{0.25}$. In disordered Ta/Co-Ni multilayer/$Pt_{0.75}Ti_{0.25}$, the disordered Ta generates little spin current, leaving the top surface of the $Pt_{0.75}Ti_{0.25}$ the only source of the spin current in the Co-Ni multilayer. (c) Optical microscope image of the Hall bar and the measurement configuration. (d) Average resistivity of the $Pt_{0.75}Ti_{0.25}$ and (d) Saturation magnetization of the [$Pt_{0.75}Ti_{0.25}$/Co-Ni multilayer/Ta]$_n$ superlattices with different repeat number $n$.

As plotted in Fig. 1(d), the average resistivity of the $Pt_{0.75}Ti_{0.25}$ layers within the [$Pt_{0.75}Ti_{0.25}$ (2 nm)/Co-Ni multilayer/Ta (1.5 nm)]$_n$ superlattices is 91 ± 5 μΩ cm as estimated from the conductance enhancement of the Ta (1 nm)/[$Pt_{0.75}Ti_{0.25}$ (2 nm)/Co-Ni multilayer/Ta (1.5 nm)]$_n$/MgO (2 nm)/Ta(1.5 nm) compared to $n$ times the conductance of a control stack without a $Pt_{0.75}Ti_{0.25}$ layer, *i.e.*, Ta (1 nm)/Co-Ni multilayer/Ta (1.5 nm)/MgO (2 nm)/Ta(1.5 nm). The saturation magnetization ($M_s$) of each Co-Ni multilayer is 526 ± 20 emu/cm$^3$ as measured by a superconducting quantum interference device (Fig. 1(e) and the Supplementary Materials). These results suggest that the electrical and magnetic properties of each repeat of the [$Pt_{0.75}Ti_{0.25}$ (2 nm)/Co-Ni multilayer/Ta (1.5 nm)]$_n$ superlattices with $n \leq 6$ are essentially the same. Note that we find it challenging to obtain such high-quality superlattices with $n$ much greater than 6 due to the degradation of the identicalness of the repeats.

The SOTs and the PMA were characterized from the harmonic Hall voltages [14,38] under a sinusoidal electric field ($E$) of 33.3 kV/m along the $x$ direction (Fig. 1(c)) in the low-field macrospin regime by taking into account thermoelectric effects [39]. To assure the macrospin behavior and avoid possible multidomain artifacts, the samples are first saturated to +$M_z$ (-$M_z$) state by a large positive (negative) out-of-plane magnetic field ($H_z$). The in-phase first harmonic Hall voltage ($V_{1\omega}$) and the out-of-phase second harmonic Hall voltage ($V_{2\omega}$) are then collected using a low-noise lock-in amplifier while the magnetic field is swept along different directions in the low field range. As representatively shown in Fig. 2(a), the superlattice samples exhibit fairly square first harmonic Hall voltage hysteresis loops as a function of swept perpendicular magnetic field ($H_z$), revealing the good PMA of the superlattices. As shown in Fig. 2(b), $V_{1\omega}$ is a parabolic function of the in-plane magnetic field, but $V_{2\omega}$ scales linearly with the in-plane magnetic field. More data can be found in the Supplementary Materials.

The square $V_{1\omega}$ - $H_z$ loops, the parabolic in-plane magnetic field dependence of $V_{1\omega}$, and the linear in-plane magnetic field dependence of $V_{2\omega}$ consistently reveal that the superlattices in this work obey the macrospin model in the low-field regions [1], which allows accurate estimation of the SOTs from the harmonic Hall voltages. The dampinglike and fieldlike SOT fields, $H_{DL}$ and $H_{FL}$, can be estimated as [39]

$$H_{DL} = -2\frac{\partial V_{2\omega}}{\partial H_x} / \frac{\partial^2 V_{1\omega}}{\partial^2 H_x} - 2H_k V_{ANE,z}/V_{AHE}, \quad (1)$$

$$H_{FL} = -2\frac{\partial V_{2\omega}}{\partial H_y} / \frac{\partial^2 V_{1\omega}}{\partial^2 H_y} - H_{Oe}, \quad (2)$$

where $V_{AHE}$ is the anomalous Hall voltage that can be determined from the $V_{1\omega}$ - $H_z$ loops, $H_k$ is the PMA field that can be estimated from the parabolic in-plane magnetic field dependence of $V_{1\omega}$ following the relation $V_{1\omega} \approx V_{AHE}(1 - H_x^2/2H_k^2)$. $H_{Oe}$ is the transverse Oersted field [30,31] on the Co-Ni multilayers by the in-plane charge current in other layers and estimated following the same method in Ref. 23. The anomalous Nernst voltage, $V_{ANE,z}$, due to out-of-plane thermal gradient can be measured from $V_{2\omega}$ when the magnetization is aligned to the $x$ direction by applying a sufficiently large longitudinal magnetic field (Fig. 2(c)).

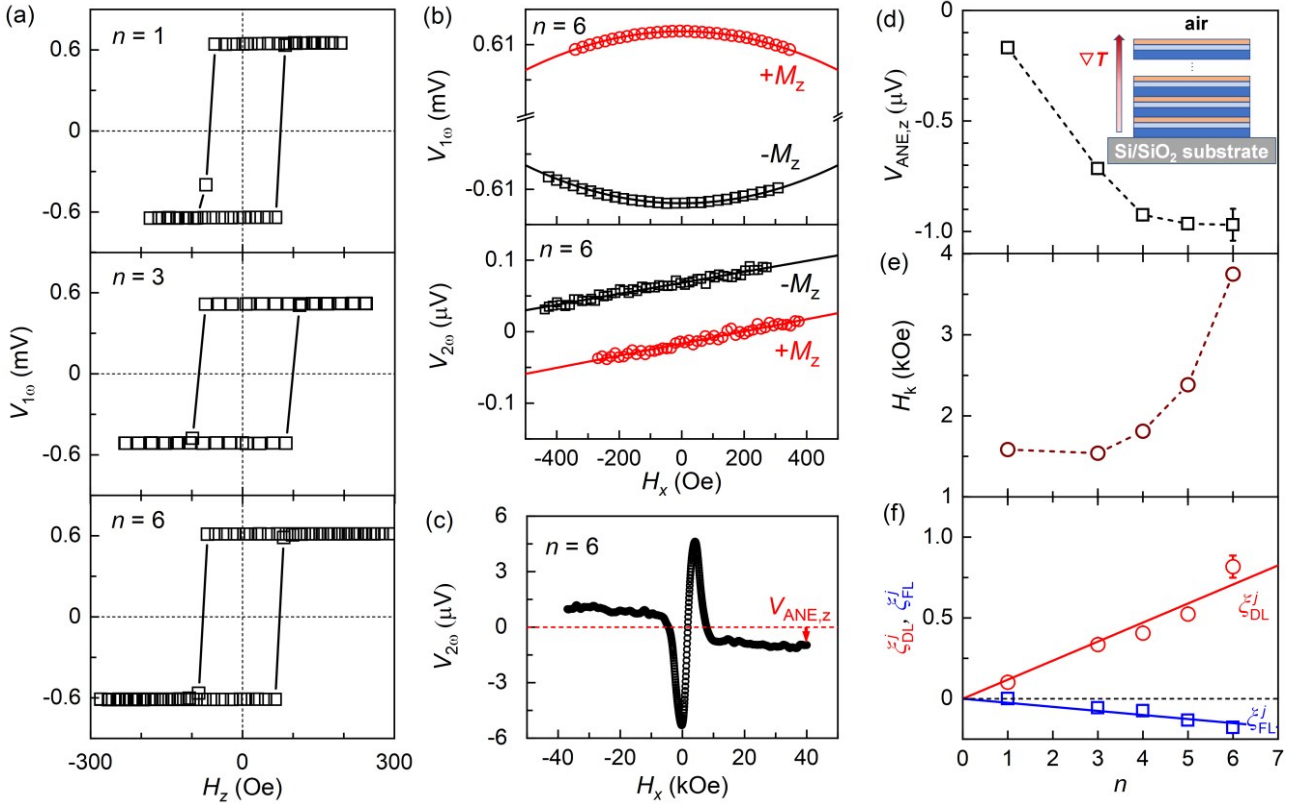

Fig. 2. Harmonic Hall voltage measurements on the [Pt$_{0.75}$Ti$_{0.25}$/Co-Ni multilayer/Ta]$_n$ superlattices. (a) First harmonic Hall voltage ($V_{1\omega}$) vs the perpendicular magnetic field for the superlattices with $n = 1$, 3, and 6. (b) First ($V_{1\omega}$) and second ($V_{2\omega}$) harmonic Hall voltages plotted as a function of in-plane field $H_x$, for the superlattice with $n = 6$. (c) $V_{2\omega}$ vs $H_x$ in the large field range for the determination of the anomalous Nernst voltage ($V_{ANE,z}$) for the superlattice $n = 6$. Variation with $n$ of (d) Anomalous Nernst voltage ($V_{ANE,z}$), (e) Perpendicular magnetic anisotropy field ($H_k$), and (f) Dampinglike and fieldlike SOT efficiencies ($\xi_{DL}^j$) and SOT efficiency ($\xi_{FL}^j$). The inset in (d) schematically depicts the formation of the out-of-plane thermal gradient due to distinct thermal dissipation rates at the top and bottom interfaces that are in contact with the air and the substrate. The solid lines in (f) denote the linear fits of the data.

As summarized in Fig. 2(d), $V_{ANE,z}$ is negative in sign and increases in magnitude as the repetition number of the superlattice increases, which is consistent with those in the [Pt/Co/MgO]$_n$ superlattices [22], the [PtCu/Co/Ta]$_n$ superlattices [23], and the [Pt/Co/W]$_n$ superlattices [40]. Microscopically, $V_{ANE,z}$ of a magnetic layer is determined by the product of the width of device (fixed at 5 μm in this work) and the average electric field generated by the anomalous Nernst effect, $\boldsymbol{E}_{ANE} = S_{xy}\boldsymbol{\nabla}T_{FM} \times \boldsymbol{M}/|\boldsymbol{M}|$, where $S_{xy}$, $\boldsymbol{\nabla}T_{FM}$, $\boldsymbol{M}$ are the anomalous Nernst coefficient, the out-of-plane thermal gradient, and magnetization vector of the magnetic layer, respectively. As schematically depicted in the inset of Fig. 2(d), the out-of-plane thermal gradient arises due to the different thermal dissipation at the top and bottom surfaces that are in contact with the air and substrate, respectively. As suggested by the finite-element analysis [23], the increase of $V_{ANE,z}$ with $n$ in such superlattices is most likely due to the increased thermal gradient $\boldsymbol{\nabla}T_{FM}$ at large total thicknesses, while the anomalous Nernst coefficient $S_{xy}$ is expected to vary little with $n$. Despite that $S_{xy}$ was suggested to increase with the density of the interfaces in some [Pt/Fe]$_n$ multilayers and [Pt/Ni]$_n$ multilayers [41,42], the [Pt$_{0.75}$Ti$_{0.25}$/Co-Ni multilayer/Ta]$_n$ superlattices have the same density of the interfaces in each repeat and thus should have the same $S_{xy}$.

The giant unintentionally-generated anomalous Nernst voltage indicates that it is crucial to take into account $V_{ANE,z}$ in the harmonic Hall voltage analysis of SOTs in superlattices and multilayers, especially when $V_{ANE,z}$ is significant or even comparable to or greater than the harmonic Hall voltage signals of the SOTs. As we have demonstrated previously [22], neglect of the strong anomalous Nernst effect, the dominant source of the second harmonic Hall voltage signals, in superlattices and multilayers leads to erroneous, giant estimates for the dampinglike SOT efficiencies in multilayers such as effectively symmetry-preserved [Pt/Co]$_n$ whose actual SOTs are negligibly small. More interestingly, $H_k$ is considerably enhanced by increasing the repeat number $n$ (Fig. 2(e)), which is highly desirable for the development of nanoscale memory, computing, and sensor devices with enhanced thermal stability and long data retention.

The efficiencies of the dampinglike SOT ($\xi_{DL}^j$) and fieldlike SOT ($\xi_{FL}^j$) are estimated as [43]

$$\xi_{DL(FL)}^j = \left(\frac{2e}{\hbar}\right)\mu_0 M_s t_{FM} H_{DL(FL)} \rho_{xx} / E, \quad (3)$$

where $e$ is the elementary charge, $\hbar$ is the reduced Planck's constant, $\mu_0$ is the permeability of vacuum, $t_{FM}$ is the total thickness of the magnetic layer ($t_{FM} = 1.95n$ nm), and $\rho_{xx}$ is the resistivity of the spin-current generator Pt$_{0.75}$Ti$_{0.25}$ (i.e., $\rho_{PtTi}$ in Fig. 1(d)). As summarized in Fig. 2(f), both $\xi_{DL}^j$ and $\xi_{FL}^j$ of the superlattice samples exhibit a linear increase with the repeat number $n$, i.e., $\xi_{DL}^{j} \approx 0.12n$ and $\xi_{FL}^{j} \approx 0.0025n$. This indicates that every repetition of the superlattice contributes nearly identical amounts of dampinglike SOT (0.10) and fieldlike SOT (0.0025). We stress that since the SOT efficiencies are defined as the magnitude of the SOTs under unit current density, the linear-in-$n$ enhancement of the SOT efficiencies of such symmetry-broken spin-orbit superlattices arises mainly from the increased total thickness of the magnetic layer but there is only minor variation expected in the SOT fields ($H_{DL}$ and $H_{FL}$) or the spin Hall ratios of the $Pt_{0.75}Ti_{0.25}$ layers during the stacking of the superlattices. The linear-in-$n$ enhancement of the SOTs and the fieldlike SOT being of the opposite sign and much smaller in magnitude than the dampinglike SOT agree well with the SHE being the dominant origin of the SOTs in the symmetry-broken spin-orbit superlattices and reveal negligible stacking-collectively-induced contribution of SOTs or spin currents. This is consistent with the linear increase of the dampinglike and fieldlike SOTs with $n$ in the in-plane anisotropy $[Pt_{0.75}Cu_{0.25}/Co/Ta]_n$ superlattices [22] and PMA $[Pt/Co/W]_n$ superlattices [40] and the absence of SOTs in the symmetry-perservered $[Pt/Co]_n$ superlattices [22].

The spin Hall spin current responsible for the linear-in-$n$ SOTs in $[Pt_{0.75}Ti_{0.25}/Co$-Ni multilayer/Ta$]_n$ superlattice should be attributed to the SHE of the $Pt_{0.75}Ti_{0.25}$ since the experiments of ours and other groups [13-23] have consistently indicated negligible SOTs in Ta 1.5/FM bilayers. The latter is likely related to the disordered structure in the Ta due to the low sputter-deposition rate [37]. Note that even the $\beta$-phased Ta with a large thickness of 5 nm has been reported to generate a very small SOT efficiency of -0.01 in some $Ta/Ni_{81}Fe_{19}$ heterostructures [44,45], let alone the highly disordered Ta with much smaller thickness and much weaker spin Hall effect. We also note that any nonzero spin current generated from the 1.5-nm Ta would add to rather than subtract from the spin current generated from the bottom $Pt_{0.75}Ti_{0.25}$ within the same superlattice repeat (Fig. 1(b)) because of the opposite signs of the spin Hall conductivities for Ta [2] and $Pt_{0.75}Ti_{0.25}$ [24], ultimately leading to further increase of the SOTs. The linear-in-$n$ scaling of $\xi_{DL}^{j}$ also suggests that each of the 1.5 nm Ta symmetry-breaking layers has played an ideal role in decoupling the spin currents of different superlattice repeats by blocking the detrimental flow of the spin current from the $Pt_{0.75}Ti_{0.25}$ overlayer within the neighboring superlattice repeat towards the Co-Ni multilayer within the same repeat of the Ta.

The $\xi_{DL}^{j}/n$ value of 0.12 for the $[Pt_{0.75}Ti_{0.25}/Co$-Ni multilayer/Ta$]_n$ superlattice is smaller than that of the $Pt_{0.75}Ti_{0.25}$ 5.6/Fe 6 bilayer in our previous work, which can be attributed to the small layer thickness of the $Pt_{0.75}Ti_{0.25}$ layer in the superlattice and enhanced spin memory loss at the $Pt_{0.75}Ti_{0.25}/Co$-Ni and Co/Ni interfaces. As discussed previously [46], post-annealing can enhance the interfacial PMA as well as spin-memory loss via increased interfacial spin-orbit scattering of spins into the atomic lattice. However, we do not attribute the low SOTs to any low-temperature thermal annealing-induced atomic intermixing because atomic intermixing at the interfaces has been indicated to lower the interfacial spin-orbit coupling and thus to enhance the SOTs via increasing the spin transparency of HM/FM interfaces [47].

Finally, we demonstrate that the $[Pt_{0.75}Ti_{0.25}/Co$-Ni multilayer/Ta$]_n$ superlattices, despite the giant magnetic layer thickness of up to 11.7 nm, can be switched by current-induced SOTs under an in-plane assisting magnetic field ($H_x$). As we schematically show in Fig. 3(a), during the current switching measurement, short square current pulses of 67 μs in duration ($j_{pulse1}$) were sourced to the superlattice Hall bar devices as the write current using a Keithley 6221. To read the Hall resistance of the device after the each write current pulse by a Keithley 2182A voltage-meter, a long and low read current pulse ($j_{pulse2}$) of 1s in duration and of 0.1 mA in magnitude was source into the Hall-bar device with a 3 second delay relative to the write current pulse as the read excitation using another Keithley 6221. Here, we have avoided using the same current for write and read processes to reduce the influence of Joule heating on the detected Hall voltage and thus Hall resistance. The current switching hysteresis loops of the superlattices with different $n$ ($H_x$ = 100 Oe) are plotted in Fig. 3(b). As summarized in Fig. 3(c), the switching current averaged from the upward and downward switching ($j_{c,\uparrow}$-$j_{c,\downarrow}$)/2 remains approximately the same for the superlattices with different repeat number $n$ due to the reasonable identicalness of the superlattice repeats despite the increase in the thickness of the magnetic layer. As shown in Fig. 3(d), the critical current densities for the upward ($j_{c,\uparrow}$) and downward ($j_{c,\downarrow}$) switching only exhibit a weak dependence on the in-plane magnetic field, which is similar to previous reports of Co-based PMA samples (see Fig. S1 in Ref. [12] for data of Pt-X/Co bilayers and Ref. [40] for the data of $[Pt/Co/W]_n$ superlattices).

While the switching current density of $1.5 \times 10^7 A/m^2$ is within the range of those for switching only 1 nm thick 3$d$ ferromagnetic layer in conventional heavy metal/ferromagnet bilayers [14-19], we emphasize that current-driven switching of such thick magnetic layers is unexpected in conventional heavy metal/ferromagnet bilayer scheme, for which a 3$d$ magnetic layer is unlikely to be switchable by interfacial SOT before the blowout of the device when the thickness is several nm or even greater, e.g., 11.7 nm. The ability of the superlattice to effectively switch a large-volume magnetic layer is interesting for spintronic applications where the magnetic layer has to be thick to maintain a PMA and/or thermal stability when the lateral size scales down to tens of nanometers.

Since the power scales with $(1/\xi_{DL}^j)^2$, such superlattices should have $n^2$ times lower power consumption to generate a given spin torque strength than the corresponding conventional magnetic bilayer with the same total magnetic thickness. Such superlattices can be utilized to develop self-torqued magnetic tunnel junctions and superlattice racetracks for memory, computing, and microwave emitter applications that require low power, long endurance, low impedance, and large magnetic volume for enhanced thermal stability or signal output. It is interesting to note that the switching current density is not symmetric for upward and downward switching of the $n = 1$ sample (Fig. 3(d)), which, as discussed in detail in Refs. [48,49], can be attributed possibly to the antisymmetric anti-domain nucleation or chiral perpendicular magnetic field associated with the Dzyaloshinskii-Moriya interaction and magnetic non-uniformity.

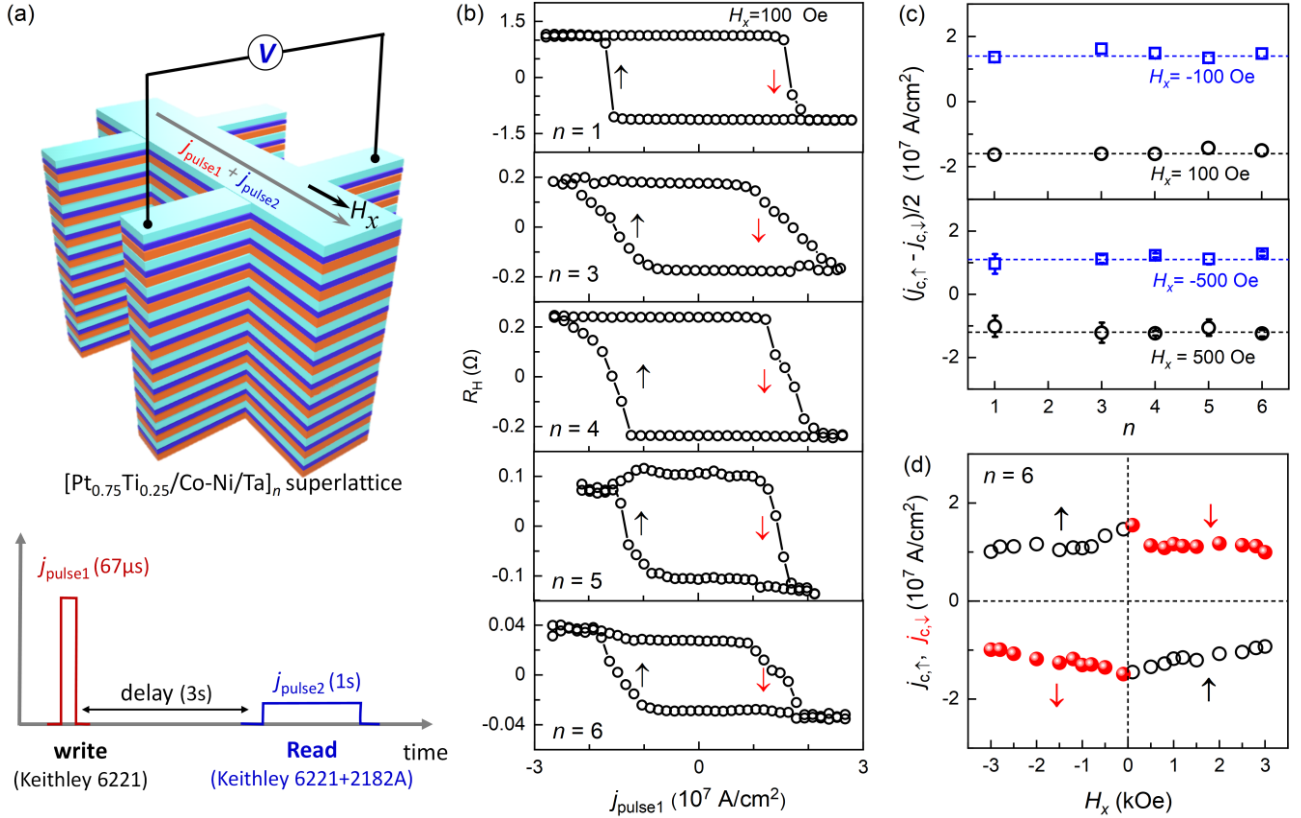

Fig. 3. Current switching of the [Pt$_{0.75}$Ti$_{0.25}$ (2 nm)/Co-Ni multilayer/Ta (1.5 nm)]$_n$ superlattices. (a) Schematic of the measurement configuration for current switching (see the main text for detailed discussion). (b) Hall resistance vs the average current density ($j_{pulse1}$) for the superlattice devices with different $n$ under a longitudinal magnetic field of $H_x = 100$ Oe, with the black and red arrows indicating the upward and downward switching. (c) Dependence on $n$ of the average switching current density, $(j_{c,\uparrow}-j_{c,\downarrow})/2$, for the superlattice devices under in-plane magnetic field of $H_x = \pm 100$ Oe and $H_x = \pm 500$ Oe. (d) Dependence on $H_x$ of the upward and downward switching current densities ($j_{c,\uparrow}$ and $j_{c,\downarrow}$) for the superlattice device with $n = 6$.

We also mention that the current switching results in Fig. 3(b) suggest Hall resistance variations that are 100%, 78%, 78%, 74% and 50% of that from $H_z$- driven switching for the superlattice with $n = 1, 3, 4, 5$, and 6, respectively. The reduced switching ratio in the thick, large-$n$ superlattice devices is likely due to partial switching. Similar partial current switching under in-plane magnetic fields has been widely observed in various ferromagnetic and ferrimagnetic PMA samples [50-56], but the exact microscopic mechanism has been a long-standing open question and beyond the scope of this letter. However, we stress that the switching current density of the partial switching for the [Pt$_{0.75}$Ti$_{0.25}$/Co-Ni multilayer/Ta]$_n$ superlattices still represents an important parameter for the potential application since such switching is well repeatable and deterministic. Such partial switching can be also technologically useful. Note that even the switching of small skyrmions within unswitched magnetic host has been already demonstrated to electrically detectable by magnetic tunnel junction resistance [57].

In summary, we have developed the [Pt$_{0.75}$Ti$_{0.25}$/Co-Ni multilayer/Ta]$_n$ superlattice with strong spin-orbit torque, perpendicular magnetic anisotropy, and efficient current switching at the same time. We demonstrate that the efficiency of the spin-orbit torque increases linearly with the repetition number $n$, which is in good agreement with the spin Hall effect being the only source of the spin-orbit torque. Meanwhile, the perpendicular magnetic anisotropy field is also enhanced by more than a factor of 2 as $n$ increases from 1 to 6. The [Pt$_{0.75}$Ti$_{0.25}$/Co-Ni multilayer/Ta]$_n$ superlattice with large thickness also allows to be further patterned into very thin nanopillars following Watanabe et al. [58] for strong perpendicular shape magnetic anisotropy and thus achieve considerably enhanced thermal stability and data retention. The [Pt$_{0.75}$Ti$_{0.25}$/Co-Ni multilayer/Ta]$_n$ superlattice also exhibits deterministic magnetization switching despite the very large layer thicknesses at relatively low current densities. The combination of the strong spin-orbit torque, perpendicular magnetic anisotropy, low impedance, and low-current switching makes [Pt$_{0.75}$Ti$_{0.25}$/Co-Ni multilayer/Ta]$_n$ superlattice a compelling material candidate for ultrafast, energy-efficient, long-data-retention spintronic memory and logic technologies.

**Acknowledgments**
This work is supported partly by the National Key Research and Development Program of China 2022YFA1204000), the Beijing National Natural Science Foundation (Z230006), and the National Natural Science Foundation of China (12274405 and 12393831).

**Reference**
1. L. Zhu, Switching of Perpendicular Magnetization by Spin–Orbit Torque. Adv. Mater. 35, 2300853 (2023).
2. L. Liu, C.-F. Pai, Y. Li, H. W. Tseng, D. C. Ralph, R. A. Buhrman, Spin-torque switching with the giant spin Hall effect of tantalum, Science 336, 555 (2012).
3. I. M. Miron, K. Garello, G. Gaudin, P.-J. Zermatten, M. V. Costache, S. Auffret, S. Bandiera, B. Rodmacq, A. Schuhl, P. Gambardella, Perpendicular switching of a single ferromagnetic layer induced by in-plane current injection. Nature (London) 476, 189 (2011).
4. Z. Luo, A. Hrabec, T. P. Dao, G. Sala, S. Finizio, J. Feng, S. Mayr, J. Raabe, P. Gambardella, L. J. Heyderman, Current-driven magnetic domain-wall logic, Nature, 579, 214-218 (2020).
5. Y. Zhang, H. Xu, K. Jia, G. Lan, Z. Huang, B. He, C. He, Q. Shao, Y. Wang, M. Zhao, T. Ma, J. Dong, C. Guo, C. Cheng, J. Feng, C. Wan, H. Wei, Y. Shi, G. Zhang, X. Han, and G. Yu, Room temperature field-free switching of perpendicular magnetization through spin-orbit torque originating from low-symmetry type II Weyl semimetal, Sci. Adv. 9, adg9819 (2023).
6. S. Hu, D. Shao, H. Yang, C. Pan, Z. Fu, M. Tang, Y. Yang, W. Fan, S. Zhou, E. Y. Tsymbal, X. Qiu, Efficient perpendicular magnetization switching by a magnetic spin Hall effect in a noncollinear antiferromagnet, Nat. Commun. 13, 4447 (2022).
7. L. Liu, C. Zhou, T. Zhao, B. Yao, J. Zhou, X. Shu, S. Chen, S. Shi, S. Xi, D. Lan, W. Lin, Q. Xie, L. Ren, Z. Luo, C. Sun, P. Yang, E.-J. Guo, Z. Dong, A. Manchon, J. Chen, Current-induced self-switching of perpendicular magnetization in CoPt single layer, Nat. Commun. 13, 3539 (2022).
8. S. Liang, A. Chen, L. Han, H. Bai, C. Chen, L. Huang, M. Ma, F. Pan, X. Zhang, C. Song, Field-Free Perpendicular Magnetic Memory Driven by Out-of-Plane Spin-Orbit Torques, Adv. Funct. Mater. 35, 2417731 (2024).
9. Y. Fan, Q. Wang, W. Wang, D. Wang, Q. Huang, Z. Wang, X. Han, Y. Chen, L. Bai, S. Yan, and Y. Tian, Robust Magnetic-Field-Free Perpendicular Magnetization Switching by Manipulating Spin Polarization Direction in RuO$_2$/[Pt/Co/Pt] Heterojunctions, ACS Nano 18, 26350 (2024).
10. R. Li, S. Zhang, S. Luo, Z. Guo, Y. Xu, J. Ouyang, M. Song, Q. Zou, L. Xi, X. Yang, J. Hong, and L. You, A spin–orbit torque device for sensing three-dimensional magnetic fields. Nat. Electron. 4, 179 (2021).
11. O. J. Lee, L. Q. Liu, C. F. Pai, Y. Li, H. W. Tseng, P. G. Gowtham, J. P. Park, D. C. Ralph, R. A. Buhrman, Central role of domain wall depinning for perpendicular magnetization switching driven by spin torque from the spin Hall effect, Phys. Rev. B, 89, 024418 (2014).
12. L. Zhu, D.C. Ralph, R.A. Buhrman, Lack of Simple Correlation between Switching Current Density and Spin-Orbit-Torque Efficiency of Perpendicularly Magnetized Spin-Current-Generator-Ferromagnet Heterostructures, Phys. Rev. Appl. 15, 024059 (2021).
13. L. Zhu, D. C. Ralph, and R. A. Buhrman, Maximizing spin-orbit torque generated by the spin Hall effect of Pt. Appl. Phys. Rev. 8, 031308 (2021).
14. L. Zhu, D. C. Ralph, and R. A. Buhrman, Highly Efficient Spin-Current Generation by the Spin Hall Effect in Au$_{1-x}$Pt$_x$. Phys. Rev. Appl. 10, 031001 (2018).
15. L. Zhu, L. Zhu, M. Sui, D. C. Ralph, and R. A. Buhrman, Variation of the giant intrinsic spin Hall conductivity of Pt with carrier lifetime. Sci. Adv. 5, eaav8025 (2019).
16. L. Zhu, K. Sobotkiewich, X. Ma, X. Li, D. C. Ralph, and R. A. Buhrman, Strong Damping-Like Spin-Orbit Torque and Tunable Dzyaloshinskii–Moriya Interaction Generated by Low-Resistivity Pd$_{1-x}$Pt$_x$ Alloys. Adv. Funct. Mater. 29, 1805822 (2019).
17. C. Y. Hu and C. F. Pai, Benchmarking of Spin–Orbit Torque Switching Efficiency in Pt Alloys. Adv. Quantum Technol. 3, 2000024 (2020).
18. Q. Liu, J. Li, L. Zhu, X. Lin, X. Xie, L. Zhu, Strong Spin-Orbit Torque Induced by the Intrinsic Spin Hall Effect in Cr$_{1-x}$Pt$_x$, Phys. Rev. Appl. 18, 054079 (2022).
19. J. Quan, X. Zhao, W. Liu, L. Liu, Y. Song, Y. Li, X. Zhao, and Z. Zhang. Enhancement of spin–orbit torque and modulation of Dzyaloshinskii-Moriya interaction in Pt$_{100-x}$Cr$_x$/Co/AlO$_x$ trilayer. Appl. Phys. Lett. 117, 222405 (2020).
20. L. Zhu, L. Zhu, S. Shi, M. Sui, D. C. Ralph, and R. A. Buhrman, Enhancing Spin-Orbit Torque by Strong Interfacial Scattering From Ultrathin Insertion Layers. Phys. Rev. Appl. 11, 061004 (2019).
21. L. J. Zhu and R. A. Buhrman, Maximizing Spin-Orbit-Torque Efficiency of Pt/Ti Multilayers: Trade-Off Between Intrinsic Spin Hall Conductivity and Carrier Lifetime. Phys. Rev. Appl. 12, 6, 051002 (2019).
22. L. Zhu, J. Li, L. Zhu, and X. Xie, Boosting Spin-Orbit-Torque Efficiency in Spin-Current-Generator/Magnet/Oxide Superlattices, Phys. Rev. Appl. 18, 064052 (2022).
23. X. Lin, L. Zhu, Q. Liu, and L. Zhu, Giant, Linearly Increasing Spin–Orbit Torque Efficiency in Symmetry-Broken Spin–Orbit Torque Superlattices. Nano Lett. 23, 9420 (2023).

24. L. Zhu, D.C. Ralph, Strong variation of spin-orbit torques with relative spin relaxation rates in ferrimagnets, Nat. Commun. 14, 1778 (2023).
25. T. Seki, J. Shimada, S. Iihama, M. Tsujikawa, T. Koganezawa, A. Shioda, T. Tashiro, W. Zhou, S. Mizukami, M. Shirai, and K. Takanashi, Magnetic Anisotropy and Damping for Monolayer-Controlled Co|Ni Epitaxial Multilayer, J. Phys. Soc. Jpn. 86, 074710 (2017).
26. S. Fukami, H. Sato, M. Yamanouchi, S. Ikeda, and H. Ohno, CoNi Films with Perpendicular Magnetic Anisotropy Prepared by Alternate Monoatomic Layer Deposition. Appl. Phys. Express 6, 073010 (2013).
27. G. H. O. Daalderop, P. J. Kelly, and F. J. A. den Broeder, Prediction and confirmation of perpendicular magnetic anisotropy in Co/Ni multilayers. Phys. Rev. Lett. 68, 682 (1992).
28. L. You, R. C. Sousa, S. Bandiera, B. Rodmacq, and B. Dieny, Co/Ni multilayers with perpendicular anisotropy for spintronic device applications. Appl. Phys. Lett. 100, 172411 (2012).
29. M. Mohanta, S. K. Parida, A. Sahoo, Z. Hussain, M. Gupta, V. R. Reddy, and V. R. R. Medicherla, Structural and magnetic properties of CoNi surface alloys. Phys. B Condens. Matter 572, 105 (2019).
30. S. Girod, M. Gottwald, S. Andrieu, S. Mangin, J. McCord, Eric E. Fullerton, J.-M. L. Beaujour, B. J. Krishnatreya, A. D. Kent, Strong perpendicular magnetic anisotropy in Ni/Co(111) single crystal superlattices, Appl. Phys. Lett. 94, 262504 (2009).
31. J.-M. L. Beaujour, W. Chen, K. Krycka, C.-C. Kao, J. Z. Sun, and A. D. Kent, Ferromagnetic resonance study of sputtered Co|Ni multilayers. Eur. Phys. J. B 59, 475 (2007).
32. S. Mangin, D. Ravelosona, J. A. Katine, M. J. Carey, B. D. Terris, and E. E. Fullerton, Current-induced magnetization reversal in nanopillars with perpendicular anisotropy. Nat. Mater. **5**, 210 (2006).
33. D. Bedau, H. Liu, J.-J. Bouzaglou, A. D. Kent, J. Z. Sun, J. A. Katine, E. E. Fullerton, and S. Mangin, Ultrafast spin-transfer switching in spin valve nanopillars with perpendicular anisotropy. Appl. Phys. Lett. 96, 022514 (2010).
34. T. Koyama, G. Yamada, H. Tanigawa, S. Kasai, N. Ohshima, S. Fukami, N. Ishiwata, Y. Nakatani, and T. Ono, Control of Domain Wall Position by Electrical Current in structured Co/Ni wire with perpendicular magnetic anisotropy. Appl. Phys. Exp. 1, 101303 (2008).
35. T. Koyama, D. Chiba, K. Ueda, K. Kondou, H. Tanigawa, S. Fukami, T. Suzuki, N. Ohshima, N. Ishiwata, Y. Nakatani, K. Kobayashi and T. Ono, Observation of the intrinsic pinning of a magnetic domain wall in a ferromagnetic nanowire. Nat. Mater. 10, 194 (2011).
36. T. Moriyama, T. J. Gudmundsen, P. Y. Huang, L. Liu, D. A. Muller, D. C. Ralph, and R. A. Buhrman, Tunnel magnetoresistance and spin torque switching in MgO-based magnetic tunnel junctions with a Co/Ni multilayer electrode. Appl. Phys. Lett. 97, 072513 (2010).
37. R. Bansal, N. Behera, A. Kumar, and P. K. Mudulia, Crystalline phase dependent spin current efficiency in sputtered Ta thin films, Appl. Phys. Lett. 110, 202402 (2017).
38. M. Hayashi, J. Kim, M. Yamanouchi, and H. Ohno, Quantitative characterization of the spin-orbit torque using harmonic Hall voltage measurements. Phys. Rev. B 89, 144425 (2014).
39. Q. Liu, L. Zhu, X. S. Zhang, D.A. Muller, D.C. Ralph, Giant bulk spin–orbit torque and efficient electrical switching in single ferrimagnetic FeTb layers with strong perpendicular magnetic anisotropy, Appl. Phys. Rev. 9, 021402 (2022).
40. Z. Yan, Z. Li, L. Zhu, X. Lin, L. Zhu, Linear Enhancement of Spin-Orbit Torque and Absence of Bulk Rashba Spin Splitting in Perpendicularly Magnetized [Pt/Co/W]$_n$ Superlattices, Chin. Phys. Lett. 42, 090701 (2025).
41. T. Seki, Y. Sakuraba, K. Masuda, A.Miura, M. Tsujikawa, K.Uchida, T. Kubota, Y. Miura, M.Shirai, K.Takanashi, Enhancement of the anomalous Nernst effect in Ni/Pt superlattices. Phys. Rev. B 103, L020402 (2021).
42. K.-i.Uchida, T. Kikkawa, T. Seki, T. Oyake, J. Shiomi, Z. Qiu, K. Takanashi, E. Saitoh, Enhancement of anomalous Nernst effects in metallic multilayers free from proximity-induced magnetism. Phys. Rev. B 92, 094414 (2015).
43. M.-H. Nguyen, D. C. Ralph, and R. A. Buhrman, Spin Torque Study of the Spin Hall Conductivity and Spin Diffusion Length in Platinum Thin Films with Varying Resistivity. Phys. Rev. Lett. 116, 126601 (2016).
44. Q. Liu, X. Lin, Z. Nie, G. Yu, L. Zhu, Efficient generation of out-of-plane polarized spin current in polycrystalline heavy metal devices with broken electric symmetries, Adv. Mater. 36, 2406552 (2024).
45. T.-Y. Chen, Y. Ou, T.-Y. Tsai, R. A. Buhrman, C.-F. Pai, Spin-orbit torques acting upon a perpendicularly magnetized Py layer, APL Mater. 6, 121101 (2018).
46. L. Zhu, D. C. Ralph, R. A. Buhrman, Spin-orbit torques in heavy-metal-ferromagnet bilayers with varying strengths of interfacial spin-orbit coupling, Phys. Rev. Lett. 122, 077201 (2019).
47. L. Zhu, D. C. Ralph, R. A. Buhrman, Enhancement of spin transparency by interfacial alloying, Phys. Rev. B 99, 180404(R)(2019).
48. Q. Liu, L. Liu, G. Xing, and L. Zhu, Asymmetric magnetization switching and programmable complete Boolean logic enabled by long-range intralayer Dzyaloshinskii-Moriya interaction. Nat. Commun. 15, 2978 (2024).
49. G. Han, X. Lin, Q. Liu, G. Gong, L. Zhu, Invalidation Of the Domain Wall Depinning Model and Current-Induced Switching Angle Shift Analysis In Pt$_{75}$Ti$_{25}$/Ti/Fe$_{60}$Co$_{20}$B$_{20}$ Heterostructure, Adv. Mater. (2025).
50. C. O. Avci, A. Quindeau, C.-F. Pai, M. Mann, L. Caretta, A. S. Tang, M. C. Onbasli, C. A. Ross, G. S. D. Beach, Current-induced switching in a magnetic insulator,

Nat. Mater. 16, 309–314 (2017).

51. Z. Zhao, M. Jamali, A. K. Smith, J.-P. Wang, Spin Hall switching of the magnetization in Ta/TbFeCo structures with bulk perpendicular anisotropy, Appl. Phys. Lett. 106, 132404 (2015).
52. R. Yoshimi et al., Current-driven magnetization switching in ferromagnetic bulk Rashba semiconductor (Ge,Mn)Te, Sci. Adv. 4, eaat9989 (2018).
53. L. Liu, Q. Qin, W. Lin, C. Li, Q. Xie, S. He, X. Shu, C. Zhou, Z. Lim, J. Yu, W. Lu, M. Li, X. Yan, S. J. Pennycook, J. Chen, Current-induced magnetization switching in all-oxide heterostructures, Nat. Nanotechnol. 14, 939–944 (2019).
54. Q. Huang, Y. Dong, X. Zhao, J. Wang, Y. Chen, L. Bai, Y. Dai, Y. Dai, S. Yan, and Y. Tian, Electrical Control of Perpendicular Magnetic Anisotropy and Spin-Orbit Torque-Induced Magnetization Switching, Adv. Electron. Mater. 6, 1900782 (2020).
55. S. N. Kajale, T. Nguyen, N. T. Hung, M. Li, D. Sarkar, Field-free deterministic switching of all–van der Waals spin-orbit torque system above room temperature, Sci. Adv. 10, eadk8669 (2024).
56. Y. Zhang, X. Ren, R. Liu, Z. Chen, X. Wu, J. Pang, W. Wang, G. Lan, K. Watanabe, T. Taniguchi, Y. Shi, G. Yu, Q. Shao, Robust Field-Free Switching Using Large Unconventional Spin-Orbit Torque in an All-Van der Waals Heterostructure, Adv. Mater. 36, 2406464 (2024).
57. Y. Guang, L. Zhang, J. Zhang, Y. Wang, Y. Zhao, R. Tomasello, S. Zhang, B. He, J. Li, Y. Liu, J. Feng, H. Wei, M. Carpentieri, Z. Hou, J. Liu, Y. Peng, Z. Zeng, G. Finocchio, X. Zhang, J. M. D. Coey, X. Han, G. Yu, Electrical Detection of Magnetic Skyrmions in a Magnetic Tunnel Junction, Adv. Electron. Mater. 9, 2200570 (2023).
58. K. Watanabe, B. Jinnai, S. Fukami, H. Sato, and H. Ohno, Shape anisotropy revisited in single-digit nanometer magnetic tunnel junctions. Nat. Commun. 9, 663 (2018).